\documentstyle[12pt]{article}
\begin{document}
\begin{titlepage}
\noindent
May, 1992\hfill{UM-P-92/27}

\noindent
Revised July, 1992\hfill{OZ-92/7}

\hfill{BA-36}

\begin{center}
{\Large {\bf Bounds on Mini-charged Neutrinos in the\\
Minimal Standard Model}}\\
\vspace{7 mm}
K. S. Babu$^{(a)}$ and R. R. Volkas$^{(b)}$\\
\vspace{3 mm}
\end{center}
{\it $^{(a)}$Bartol Research Institute, University of Delaware,
Newark, DE 19716 U.S.A.}
\vspace{1 mm}

\noindent
{\it $^{(b)}$Research Centre for High Energy Physics, School of Physics,
University of Melbourne, Parkville 3052 Australia}

\vspace{8 mm}
\centerline{ABSTRACT}
\vspace{2 mm}

In the minimal Standard Model (MSM) with three generations of quarks
and leptons, neutrinos can have tiny charges consistent with
electromagnetic gauge invariance. There are three types of
non-standard electric charge, given by $Q_{st} +
\epsilon(L_i - L_j)$, where $i, j = e, \mu, \tau$ $(i \neq j)$,
$Q_{st}$ is standard electric charge, $L_i$ is a family-lepton--number,
and $\epsilon$ is an arbitrary parameter which is put equal
to zero in the usual incarnation of the MSM. These three non-standard
electric charges are of considerable theoretical interest because
they are compatible
with the MSM Lagrangian and $SU(3)_c \otimes SU(2)_L \otimes U(1)_Y$
gauge anomaly cancellation. The two most conspicuous implications
of such non-standard electric charges are the presence
of two generations of massless charged neutrinos and a breakdown
in electromagnetic universality for $e$, $\mu$ and $\tau$.
We use results from (i) charge conservation in $\beta$-decay, (ii)
physical consequences of charged atoms in various contexts,
(iii) the anomalous
magnetic moments of charged leptons, (iv) neutrino-electron scattering,
(v) energy loss in red giant and white dwarf stars, and (vi)
limits on a cosmologically induced thermal photon mass,
to place bounds on $\epsilon$. While the constraints derived for
$\epsilon$ are rather severe in the $L_e-L_{\mu,\tau}$ cases ($\epsilon <
10^{-17}-10^{-21}$), the $L_{\mu}-L_{\tau}$ case allows $\epsilon$ to be as
large as about $10^{-14}$.
\end{titlepage}

The study of electromagnetism is one of the most fundamental activities
of both theoretical and experimental physics. In the relativistic quantum
domain germane to particle physics, electromagnetism is very
successfully described through the direct coupling of massless photons
to electrically charged particles via the familiar vector current
interaction. In the minimal
Standard Model (MSM), one genus of fermion -- the neutrino -- is taken
to have no direct coupling with photons. However, it is not actually
mandatory within the structure of the MSM for neutrinos to possess exactly
zero electric charge. The purpose of this paper is to investigate the
dramatic consequences of not having neutrinos with precisely zero electric
charge in the MSM.

In order to understand how charged neutrinos can arise in the MSM, it is
necessary to study the global symmetries of the theory.
The MSM exhibits five $U(1)$ invariances which
commute with its non-Abelian gauge symmetry group $SU(3)_c \otimes SU(2)_L$.
One of these is the Abelian gauge symmetry $U(1)_Y$ where $Y$ is the
generator of weak hypercharge, while the other four are the
symmetries $U(1)_B$ and $U(1)_{L_{e,\mu,\tau}}$, where $B$ and $L_{e,\mu,\tau}$
are baryon number and the family-lepton--numbers respectively. The usual
version of the MSM is constructed so that
these last four groups are automatic global symmetries of the classical
Lagrangian, having no associated gauge fields.

An interesting, non-trivial constraint on gauge models is anomaly
cancellation. This is often imposed so that the standard proof of the
renormalizability of gauge theories applies. Alternatively, one may
simply demand as an aesthetic principle that quantum effects not spoil
the naive gauge invariance of a model, leading also to gauge anomaly
cancellation. However one motivates it, it is striking that in the MSM
all gauge anomalies from $SU(3)_c \otimes SU(2)_L \otimes U(1)_Y$ cancel
within each fermion family. In model-building one usually finds that
anomaly-cancellation imposes severe constraints on the allowed $U(1)$ charges.

It is interesting to note, therefore, that $U(1)_Y$ is not the only
anomaly-free Abelian invariance of the MSM. A simple calculation
demonstrates that differences in family-lepton--numbers are also
completely anomaly-free\footnote{Note that the cancellation of the mixed
gauge-gravitational anomaly is required in order to derive
these invariances as the unique anomaly-free set.}
with respect to $SU(3)_c \otimes SU(2)_L \otimes
U(1)_Y$. For a three family model, there are thus three of these
anomaly-free combinations, given by
\begin{equation}
L_{e\mu} \equiv L_e - L_{\mu}\quad {\rm or}\quad L_{e\tau} \equiv L_e -
L_{\tau}\quad {\rm or}\quad L_{\mu\tau} \equiv L_{\mu} - L_{\tau}.
\label{lelmu}
\end{equation}
It is important to understand that although each of these differences is
individually anomaly-free, no two are anomaly-free with respect to each
other.\footnote{Note that the quark analogues of Eq.~(\ref{lelmu}) are
explicitly broken in the MSM Lagrangian (this is manifested through a
non-diagonal Kobayashi-Maskawa matrix), so the only anomaly-free Abelian
invariance acting on quarks is $U(1)_Y$.}

This interesting observation immediately leads one to ask whether or not
these particular subsets of the global symmetries of the MSM have
associated gauge fields.
A possibility is that one of $U(1)_{L_{e\mu}}$,
$U(1)_{L_{e\tau}}$ or
$U(1)_{L_{\mu\tau}}$ is gauged as a local symmetry that has no role to play
in electroweak physics. The $Z'$ model which ensues has recently been
studied in the literature \cite{zprime}.

Another fascinating possibility, which will be the focus of this paper,
is for the definition of the weak hypercharge in the MSM to be altered
in one of three ways:
\begin{equation}
Y_{e\mu} = Y_{st} + 2\epsilon L_{e\mu}\quad {\rm or}\quad Y_{e\tau} =
Y_{st} + 2\epsilon L_{e\tau}\quad {\rm or}\quad Y_{\mu\tau} =
Y_{st} + 2\epsilon L_{\mu\tau},
\label{lelmuY}
\end{equation}
where $Y_{st}$ is the standard hypercharge of the MSM and $\epsilon$ is
a free parameter. After electroweak symmetry breaking, this leads to
non-standard unbroken electric charges given by
\begin{equation}
Q_{ij} = Q_{st} + \epsilon L_{ij},
\label{lelmuQ}
\end{equation}
where $Q_{st} = I_3 + Y_{st}/2$ is standard electric charge and $i,j = e, \mu,
\tau$ $(i \neq j)$. Equation (\ref{lelmuQ}) defines the precise ways in
which electric charge quantization can fail in the multi-family
MSM\footnote{If right-handed neutrinos are added to the MSM fermion
spectrum, and only Dirac neutrino masses are induced after electroweak
symmetry breaking, then the family-lepton--numbers are in general
explicitly broken and the above form of electric charge dequantization
is excluded. In this case, however, $B-L$ generates an anomaly-free
$U(1)$ symmetry, and so charge dequantization can ensue through $Q = Q_{st}
+ \epsilon(B-L)$\cite{FJLV,Desh} (for bounds on $\epsilon$ in this model
see Ref.\cite{FJLV}). If bare Majorana masses are included for the
right-handed neutrinos, then $B-L$ is also explicitly broken, and no
electric charge dequantization at all is allowed \cite{babu}.}\cite{foot}.
Note that electromagnetic gauge invariance is still exact, and the
photon as usual has no zero-temperature mass (thermal masses will be
considered later).

The electric charge generators $Q_{ij}$ alter the physical electric
charges for two out of the three families of leptons. For
instance under $Q_{\mu\tau}$,
\begin{eqnarray}
&Q_e = -1,\quad Q_{\mu} = -1 + \epsilon,\quad Q_{\tau} = -1 -
\epsilon,&\\ \nonumber
&Q_{\nu_e} = 0,\quad Q_{\nu_{\mu}} = \epsilon,\quad Q_{\nu_{\tau}} =
- \epsilon,&
\end{eqnarray}
while the quark charges assume their standard values, of course. The two
observable consequences of this are that $e$, $\mu$ and $\tau$ do not
have identical charges, and two neutrino flavors have equal and opposite
charges. The purpose of this paper is to derive phenomenological bounds
on $\epsilon$ for each of the three non-standard MSM's.\footnote{While
this paper was being written up, we came across a preprint
(Reference~\cite{tanaka1}) which quotes some bounds on charge dequantization
in the MSM, but it is our intention to do a much more thorough analysis
here.} Our
phenomenological constraints come either from physics which would be
sensitive to (small) violations of electromagnetic universality for $e$,
$\mu$ and $\tau$, or from limits connected with the
existence of mini-charged massless neutrinos.

Several phenomenological analyses on mini-charged particles have recently
been published \cite{FJLV,tanaka1,tanaka2,ksbrnm,MR,DCB,DI}.
Reference~\cite{FJLV} deals
specifically with another form of electric charge dequantization
featuring electrically charged neutrinos (see footnote 4 above),
while Ref.~\cite{tanaka1} is discussed in footnote 3 above.
The papers in Refs.~\cite{tanaka2,ksbrnm} deal with mini-charged particles in
models where electric-charge conservation is violated, while
Refs.~\cite{MR,DCB,DI}, on the other hand, examine constraints on
completely new and exotic particles of tiny electric charge. Some of the
constraints derived in these papers are immediately applicable to the
models of charge dequantisation considered here, while others are
irrelevant. It is important to determine the specific phenomenological
constraints on the parameter $\epsilon$ in Eq.~(\ref{lelmuQ}) because of
the strong theoretical underpinning it has from the structure of the
MSM.

The parameter $\epsilon$ for the $U(1)_{Y_{e\mu}}$ and
$U(1)_{Y_{e\tau}}$ cases (see Eq. (2)) can be directly
and severely constrained from
a variety of experiments. By assuming electric charge conservation
(which is exact in the models under consideration) in $\beta$-decay,
Zorn et al\cite{zorn} were able to constrain the charge of the
electron-neutrino, which in our notation leads to $|\epsilon| <
4 \times 10^{-17}$. A bound of $|\epsilon| < 10^{-19}$
is obtained from the observation of electron-neutrinos from
supernova 1987A\cite{cocconi}.
Also, since electrons now have a charge of $Q_e = -1 +
\epsilon$, atoms are no longer electrically neutral
(which is a classic signature of electric
charge dequantization). Reference \cite{marinelli} provides a
useful summary of experiments on the neutrality of matter
performed to date.
These authors obtain a bound on the electron-proton charge magnitude
difference, which translates into $|\epsilon| < 1.6 \times 10^{-21}$.

Some interesting terrestrial effects are possible if $\epsilon$ is
nonzero because the earth may be charged. We will assume first of all
that the number of protons in the earth
is equal to the number of electrons.
It is certainly possible for this assumption to be wrong, and we will
comment on this issue again a little later on.

If $\epsilon \neq 0$, then atoms are charged, and so
mutually repulsive forces will
exist between our assumed charged
earth and laboratory samples of ordinary matter. If $|\epsilon|$
is large enough, then experiments should already have been sensitive to
this. Given that no evidence of such an effect exists, we derive
upper ``bounds'' on $|\epsilon|$ below from a couple of considerations.
Note that these limits are not bounds in the rigorous sense of the word,
because our assumption that the number of protons equals the number of
electrons in the earth need not be correct.

E\"otv\"os experiments measuring the differential attraction or
repulsion of earth with samples of material A and material B
(both taken to be pure elements), lead to an
upper ``bound'' on $\epsilon$ given by
\begin{equation}
\epsilon^2 < 10^{-12} {{G m_N} \over \alpha_{em}} \Bigl({Z_A \over M_A} - {Z_B
\over M_B}\Bigr)^{-1}
\label{eotvos}
\end{equation}
where $G$ is Newton's constant, $m_N$ is the mass of a nucleon,
$\alpha_{em}$ is the electromagnetic
fine-structure constant, $Z_{A,B}$ are atomic numbers of material A and
B respectively, while $M_{A,B}$ are the masses of atoms of A and B. For
typical materials (for instance copper and lead, see Ref.~\cite{okun})
this yields
\begin{equation}
|\epsilon| < 10^{-23}
\end{equation}
or so. Although this ``bound'' is a couple of orders of magnitude better
than the limits quoted above, it should not be taken too seriously given
the electron/proton number assumption.

Experiments near the earth's surface indicate that the earth has a
radial
electric field of less than about 100 V/m \cite{ksbrnm}.
With equal proton and
electron numbers, we then obtain that
\begin{equation}
|\epsilon| < 10^{-27}.
\label{radialE}
\end{equation}
We emphasise, however, that
the assumption of equal electron and proton numbers in the earth is
important, and so this limit cannot be regarded as a rigorous bound.

How different do the electron and proton numbers need to be to
invalidate these ``bounds''? Let us examine the radial electric field
limit in more detail. A rigorous constraint can actually be derived if
the numbers of protons and electrons are allowed to vary. It is
\begin{equation}
N_p - N_e + \epsilon N_e < 10^{24}
\label{Delta}
\end{equation}
where $N_p(N_e)$ is the number of protons(electrons) in the earth. The
proton number of the earth is about $10^{51}$, so with $N_p = N_e$ we
recover the result of Eq.~(\ref{radialE}). We can ask what $\Delta
\equiv N_p - N_e$ needs to be to make the limit on $|\epsilon|$ as weak
as the bounds of $10^{-21}$ and $10^{-17}$ from atomic neutrality
and charge conservation in $\beta$-decay, respectively. Given that $N_p
\sim N_e \sim 10^{51}$, we see from Eq.~(\ref{Delta}) that
\begin{equation}
|\epsilon| \sim 10^{-21}\ \Rightarrow\ |\Delta| \sim 10^{30}\ {\rm
and}\
|\epsilon| \sim 10^{-17}\ \Rightarrow\ |\Delta| \sim 10^{34}.
\end{equation}
If we assume that nonzero $\Delta$ is due to excess electrons, then this
amounts to between $1 - 10^4$ kg of electrons. If it is due to the
presumably less mobile protons, then this is a mass in the range
$10^3-10^7$
kg. By way of comparison, a cubic metre of earth has a mass of about
5500 kg. Another way of looking at this is that it corresponds to a
number density of excess electrons or protons of about $1 - 10^4$
particles per cubic millimetre.

Note also that an interesting effect can occur at the level of galaxies.
Naively, a limit on $|\epsilon|$ may be obtained from
the observed stability of
galaxies by requiring that electrostatic repulsion not
exceed the gravitational attraction. This yields
$|\epsilon| < ({G m_N^2/10})^{1/2} = 10^{-20}$
where equal numbers of protons and electrons are again assumed. However,
the relic neutrino cloud from the Big Bang will act as a polarizable
medium at the galactic level, and so any galactic charge will be
screened to some extent. A simple order of magnitude estimate for the
screening length is $(\epsilon e T_{\nu})^{-1}$ where $T_{\nu} \simeq
2K$ is the temperature of the relic neutrinos. For $|\epsilon|$ of the
order of $10^{-20}$ the screening length is therefore expected to be
less than typical galactic radii. Therefore, galactic charges
for reasonable values of $|\epsilon|$ should be rendered unobservable.

Since all of the bounds on the $U(1)_{Y_{e\mu}}$ and
$U(1)_{Y_{e\tau}}$ models are quite severe,
the main interest of this paper is
to derive bounds on the significantly less constrained model defined by
$U(1)_{Y_{\mu\tau}}$.
We will examine several phenomenological constraints on $\epsilon$ for
this case.

The first bound is derived by
comparing the anomalous magnetic moments $a_{\mu}$ and $a_e$ of the muon
and electron, respectively. (Since the tau anomalous moment is not as
precisely measured as the other two we do not need to consider it.) The
dominant contribution which $\epsilon$ makes to the anomalous moment of
the muon comes from the 1-loop Schwinger correction, yielding
\begin{equation}
a^{(1-{\rm loop})}_{\mu} = (\epsilon - 1)^3 {\alpha_{em} \over {2\pi}}
{e \over 2m_{\mu}}
\label{amom}
\end{equation}
compared with the electron result $a^{(1-{\rm loop})}_e =
(\alpha_{em}/2\pi) (e/2m_e)$.
Keeping only linear terms in $\epsilon$ we therefore find
that the muon anomalous moment is shifted from its standard value by an
amount $\delta a_{\mu}$ given approximately by
\begin{equation}
\delta a_{\mu} \simeq - 3 \epsilon {\alpha_{em} \over 2\pi} {e \over 2m_{\mu}}.
\end{equation}
We obtain a bound by simply demanding that this shift be less than the
experimental uncertainty in $a_{\mu}$. This approach is justified
because of the impressive agreement between the measured anomalous
moments and the standard theoretical calculations. The best measurement
of $a_{\mu}$\cite{amu} has an error of
$\pm 9 \times 10^{-9} (e/2m_{\mu})$ yielding,
\begin{equation}
|\epsilon| < 10^{-6}.
\end{equation}
This bound is many orders of magnitude less than the bounds on
the gauged $U(1)_{Y_{e\mu}}$ and $U(1)_{Y_{e\tau}}$ models.
Quite apart from the specific models we are considering in this paper,
it is also interesting to note that this is the most stringent {\it
model-independent} bound on the difference in the electric charges of
the electron and muon.

The second constraint we will analyse comes from the measured
$\nu_{\mu}$-$e$ scattering cross-section $\sigma(\nu_{\mu} e)$.
When $\epsilon = 0$, this
process is well described by the exchange of a $Z^o$ gauge boson in the
$t$-channel. For nonzero $\epsilon$ there is an additional contribution
coming from $t$-channel photon exchange. We will obtain our bound by
demanding that the photon
contribution to the cross-section lie within experimental errors.

The exact expression for $\sigma(\nu_{\mu} e)$ includes direct $Z^o$,
direct photon and interference terms, and is rather complicated. The
complication arises because of the need to keep the electron mass finite
when calculating the $t$-channel photon exchange diagram. However, a
useful approximate expression is obtained by keeping only those terms
which diverge in the massless electron limit. The result for the
$\epsilon$-dependent contribution to the cross-section is,
\begin{eqnarray}
\delta\sigma(\nu_{\mu} e) & \simeq & \Bigl[{{2\pi\alpha^2_{em}} \over m^2_e}
- \Bigl({{2\pi\alpha^2_{em}} \over {m_e E_{\nu}}} -
2\sqrt{2}\alpha_{em}G_F x(1-4x)\Bigr)\ln\Bigl({E_{\nu} \over
m_e}\Bigr)\Bigr] \epsilon^2 \nonumber \\
& - & 2 \sqrt{2} \alpha_{em} G_F (1-4x) \ln\Bigl({E_{\nu} \over
m_e}\Bigr) \epsilon
\label{xsection}
\end{eqnarray}
where $x \equiv \sin^2\theta_W$, $G_F$ is the Fermi constant, $m_e$ is
the electron mass, and $E_{\nu}$ is the incident neutrino energy.

Experiments on $\nu_{\mu}$-$e$ scattering use incident neutrino energies
$E_{\nu}$ of a few GeV's\cite{nue,BBKOPST}. Therefore the
ratio $E_{\nu}/m_e$ is large
$(> 3000)$, which illustrates why the approximate cross-section of
Eq.(\ref{xsection}) is useful. By inputting the values of the various
quantities appearing in this expression, we see that the first and third
terms dominate over the second. To obtain a bound on $\epsilon$ we use
the result of the BBKOPST collaboration\cite{BBKOPST}:
\begin{equation}
\sigma(\nu_{\mu}e)/E_{\nu} = (1.85 \pm 0.25 \pm 0.27) \times
10^{-42}\ {\rm cm}^2 \ {\rm GeV}^{-1}\quad
{\rm with}\quad E_{\nu} = 1.5\ {\rm GeV}.
\label{bbkopst}
\end{equation}
By adding the statistical and systematic errors in Eq.~(\ref{bbkopst})
in quadrature, we find that
\begin{equation}
|\epsilon| < 10^{-9}
\end{equation}
with both the first and third terms in Eq.~(\ref{xsection}) of roughly
equal importance. Note that this bound is three orders of magnitude more
stringent than that from using anomalous magnetic moments.

Both of the above bounds on the gauged $U(1)_{Y_{\mu\tau}}$ version of the
MSM were derived from considerations that were purely within the ambit
of particle physics. We will now present two bounds which also require the use
of astrophysics and cosmology, and so our faith in their
veracity will be as solid or weak as our belief in the required
astrophysical and cosmological models.

It is well known that bounds on weakly-coupled
particles can be obtained by requiring that their production in stars be
not so strong as to cause premature (and unobserved) cooling. In our
case, the decay in red giant stars of massive plasmon states into charged
$\nu_{\mu}\bar{\nu}_{\mu}$ and $\nu_{\tau}\bar{\nu}_{\tau}$ pairs can
occur. These very weakly interacting neutrinos and antineutrinos can
then escape from the star, thus cooling it. The authors of
Ref.~\cite{DCB} have (effectively) calculated a bound on $\epsilon$ from
red giant cooling by demanding that the rate of energy loss per unit
volume to mini-charged neutrino-antineutrino pairs not exceed the
nuclear energy generation rate per unit volume. They estimate that the
former quantity is given by,
\begin{equation}
\Bigl({{d^2 E} \over dVdt}\Bigr)_{\nu\bar{\nu}} \simeq
10^{34} \times \epsilon^2\ {\rm ergs\ cm^3\ sec^{-1}}
\end{equation}
and requiring that this not exceed about $10^6\ {\rm ergs\ cm^3\
sec^{-1}}$ yields the bound,
\begin{equation}
|\epsilon| < 10^{-14}.
\label{redgiant}
\end{equation}
This result is interesting because it is five orders of magnitude more
stringent than the limit obtained from $\nu_{\mu}$-$e$ scattering. The
authors of Ref.~\cite{DI} were also able to derive an astrophysical
bound by looking at the cooling of white dwarf stars, obtaining
\begin{equation}
|\epsilon| < 10^{-13}
\label{whitedwarf}
\end{equation}
which is an order of magnitude less severe than Eq.~(\ref{redgiant}).
Most astrophysicists are confident that the stellar structure and
evolution of red giants and white dwarfs are sufficiently well
understood that these bounds are to be taken very seriously. It may
nevertheless be wise to caution that, due to the very nature of the subject
matter, one cannot ascribe as much confidence on these bounds as
one can on bounds of purely particle physics origin.

We now turn to an interesting cosmological consequence of having
mini-charged neutrinos. The standard hot Big Bang model of cosmology
predicts the existence of a thermal background of each flavor of
neutrino. The temperature of this bath of thermal neutrinos is found to
be slightly less than the 3K temperature of the microwave photon
background, $T_{\nu} \simeq 2$K. Because muon- and tau-neutrinos are
charged in the $U(1)_{Y_{\mu\tau}}$ model, they form a background
``cosmic plasma'' which permeates the entire universe. All particles, and
in particular photons, have to propagate through this thermal heat bath
of neutrinos. Photons will therefore acquire a nonzero ``electric mass''
from interacting with this medium (in a similar manner to the
aforementioned acquisition by photons of a nonzero plasmon
mass in stellar interiors). Known bounds
on photon electric masses\footnote{In principle, a photon can also have
a ``magnetic mass.'' However, a nonzero magnetic mass cannot arise from
thermal effects\cite{magmass}, so it is irrelevant to the present
discussion. Note
that the most stringent bounds on the photon mass are derived from
knowledge of magnetic fields, and are thus constraints on the magnetic
rather than the electric mass of the photon.}
will therefore constrain $\epsilon$, since it
is impossible for photons to avoid propagating through the neutrino
background plasma.

The thermal electric mass of the photon is calculated through the 1-loop
contribution of the charged neutrinos to the photon vacuum polarization
tensor, where the internal neutrino propagators are taken at
finite temperature. Since a similar calculation is performed in
Ref.~\cite{nieves}, we will omit technical details of how this
computation is done. The result is
\begin{equation}
[m^{el}_{\gamma}]^2 = N_{\nu} {2\pi \over 3} \epsilon^2 \alpha_{em}
(kT_{\nu})^2,
\end{equation}
where $N_{\nu} = 2$ is the number of charged neutrino flavors and $k$ is
Boltzmann's constant. The best bound on the photon electric mass comes
from a test of Gauss's Law (or, equivalently, Coulomb's Law), and
is\cite{photonmass}
\begin{equation}
m^{el}_{\gamma} < 10^{-25} GeV.
\end{equation}
The resulting bound for $\epsilon$ is therefore
\begin{equation}
\epsilon < 10^{-12}.
\end{equation}
It is interesting that this limit is stronger than those obtained from
particle physics measurements, but less severe than those obtained from
energy loss in stellar objects.

We should remark here that the derivation of the
astrophysical bound [Eqs. (17)-(18)]
and the cosmological bound [Eq. (21)] on $\epsilon$ assumes that
$\nu_{\mu}$ and $\nu_{\tau}$ have masses less than about 10 keV.
Otherwise, (a) the plasmon decay into $\nu \overline{\nu}$ will be
forbidden kinematically inside red giants and white dwarfs where the
typical temperature is of order 10 keV and (b) the
cosmological mass density constraint requires that the keV
neutrinos decay or annihiliate in the early stages of the evolution
of the universe, so that they will not be around today to give a
thermal mass to the photon.
While it is true that in the MSM the neutrinos have no
zero temperature masses (as the photon, the neutrinos also acquire
a thermal mass of order $\epsilon^2~T$ from the background photons),
by slightly modifying the Higgs sector
(e.g. adding a triplet Higgs), it is possible to give a small `Dirac'
mass for $\nu_{\mu}$ and $\nu_{\tau}$ without violating charge conservation.
The present experimental limits on the masses of $\nu_{\mu}$ and
$\nu_{\tau}$ are 270 keV and 35 MeV respectively, so it is not impossible
to invalidate the bounds in Eqs. (17)-(18) and in Eq. (21).

As noted above, by some minor modifications to the Higgs sector neutrinos
can acquire tiny masses without violating electromagnetic gauge invariance.
However, there will be no neutrino mixing and hence no neutrino oscillations
in this case. Therefore, the MSM with mini-charged neutrinos cannot account
for the apparent deficit in the flux of neutrinos coming from the sun
through any form of neutrino oscillation mechanism. On the other hand, by
utilising another class of extensions to the basic model, the neutrino
deficit may be explained by endowing $\nu_e$ with a transition magnetic
moment with either $(\nu_{\mu})^c$ or $(\nu_{\tau})^c$, depending on whether
the $U(1)_{Y_{e\mu}}$ or $U(1)_{Y_{e\tau}}$ case is considered.
This mechansim would also
have the advantage of explaining the possible anticorrelation of the solar
neutrino flux with sunspot activity.

Another important cosmological question to consider is whether charged
relic neutrinos can induce an overall charge for the universe. If they
can then electrostatic repulsion will contribute to the expansion of the
universe. The simple answer to this question is that no overall charge
for the universe will be generated because electric charge conservation
is still exact in our models. This will follow
provided, of course, that a
neutral universe is posited as an initial condition for the Big Bang.
Charged neutrinos are therefore no more problematic in this regard than
any other stable charged particles.

In summary then, we have discovered that experiments on the neutrality of
atoms places a bound given by $|\epsilon| < 10^{-21}$ on the allowed
non-standard electric charges $Q_{st} + \epsilon (L_e - L_{\mu})$ and
$Q_{st} + \epsilon (L_e - L_{\tau})$. A direct bound on the
electron-neutrino charge of
$|\epsilon| < 4 \times 10^{-17}$ is obtained from similar experiments
where charge conservation in $\beta$-decay is assumed.
However, one of the main points of our paper is that the other allowed
non-standard charge $Q_{st} + \epsilon (L_{\mu} - L_{\tau})$ is
constrained far less profoundly. Upper bounds on $\epsilon$ of $10^{-14}$,
$10^{-13}$, $10^{-12}$, $10^{-9}$ and $10^{-6}$ were derived from,
respectively, energy loss in red gaint stars, energy loss in white dwarf
stars, the thermal electric mass of the photon, $\nu_{\mu}$-$e$
scattering and the anomalous magentic moment of the muon.

\centerline{\bf Acknowledgements}

K.S.B. would like to acknowledge discussions with D. Seckel, and both
authors would like to thank him for raising the issue of a thermal
photon mass. K.S.B. is supported by the U.S. Department of Energy.
R.R.V. would like to thank M. J. Thomson for a helpful correspondence, and
X.-G. He, K. C. Hines and K. Liffman for discussions. R.R.V. is
supported by the Australian Research Council through a Queen Elizabeth
II Fellowship.


\newpage
\begin{thebibliography}{99}

\bibitem{zprime}X.-G. He, G. C. Joshi, H. Lew and R. R. Volkas,
Phys.\ Rev.\ D{\bf 43}, R22 (1991); {\it ibid.} D{\bf 44}, 2118 (1991).

\bibitem{FJLV}R. Foot, G. C. Joshi, H. Lew and R. R. Volkas, Mod.\
Phys.\ Lett.\ A{\bf 5}, 95 (1990). For reviews on electric-charge
quantization see R. Foot, G. C. Joshi, H. Lew and R. R. Volkas,
Mod.\ Phys.\ Lett.\ A{\bf 5}, 2721 (1990); R. Foot, H. Lew and
R. R. Volkas, University of Melbourne report No. UM-P-92/52.

\bibitem{Desh}N. G. Deshpande, Oregon Report OITS-107 (1979)
(unpublished).

\bibitem{babu}K. S. Babu and R. N. Mohapatra, Phys.\ Rev.\ Lett.\
{\bf 63}, 938 (1989); Phys.\ Rev.\ D{\bf 41}, 271 (1990).

\bibitem{foot}R. Foot, Mod.\ Phys.\ Lett.\ A{\bf 6}, 527 (1991). As far
as we are aware, this is the first place in which the anomly-freedom of
family-lepton--number differences was pointed out.

\bibitem{tanaka1}E. Takasugi and M. Tanaka, Osaka University preprint
OS-GE 21-91 (unpublished).

\bibitem{tanaka2}E. Takasugi and M. Tanaka, Phys.\ Rev.\ D{\bf 44}, 3706
(1991); M. Maruno, E. Takasugi and M. Tanaka, Prog.\ Theor.\ Phys.\
{\bf 86}, 907 (1991).

\bibitem{ksbrnm}K. S. Babu and R. N. Mohapatra, Phys.\ Rev.\ D{\bf 42},
3866 (1990).

\bibitem{MR}R. N. Mohapatra and I. Z. Rothstein, Phys. Lett. B{\bf 247},
593 (1990); R.N. Mohapatra and S. Nussinov, University of Maryland
Report UMD-PP-008 (1991).

\bibitem{DCB}S. Davidson, B. Campbell and D. Bailey, Phys.\ Rev.\ D{\bf
43}, 2314 (1991).

\bibitem{DI}M. I. Dobroliubov and A. Yu.\ Ignatiev, Phys.\ Rev.\ Lett.\
{\bf 65}, 679 (1990).

\bibitem{zorn}J. C. Zorn, G. E. Chamberlain and V. W. Hughes, Phys.\
Rev.\ {\bf 129}, 2566 (1963).

\bibitem{cocconi}G. Barbiellini and G. Cocconi, Nature (London)
{\bf 329}, 21 (1987).

\bibitem{marinelli}M. Marinelli and G. Morpugo, Phys.\ Lett.\
B{\bf 137}, 439 (1984). See also H. F. Dylla and J. G. King,
Phys.\ Rev.\ A{\bf 7}, 1224 (1973).

\bibitem{okun}L. B. Okun, {\it Leptons and Quarks}, pp. 166-167
(North-Holland, Amsterdam, 1984).

\bibitem{amu}J. Bailey et al., Nucl.\ Phys.\ B{\bf 150}, 1 (1979).

\bibitem{nue}For a summary of these experiments see U. Amaldi et al.,
Phys.\ Rev.\ D{\bf 36}, 1385 (1987) and G. Costa et al., Nucl.\ Phys.\
B{\bf 297}, 245 (1988).

\bibitem{BBKOPST}K. Abe et al., Phys.\ Rev.\ Lett.\ {\bf 58}, 636 (1987).

\bibitem{magmass}E. S. Fradkin, {\it 1965 Proceedings of the Lebedev Physical
Institute} {\bf 29} 7-138 (English translation 1967 by Consultants Bureau,
New York).

\bibitem{nieves}J. F. Nieves, P. B. Pal and D. G. Unger, Phys.\ Rev.\
D{\bf 28}, 908 (1983)

\bibitem{photonmass}E. R. Williams, J. E. Faller and H. A. Hill, Phys.\
Rev.\ Lett.\ {\bf 26}, 721 (1971).

\end{thebibliography}
\end{document}